\documentclass[fleqn,usenatbib]{mnras}

\usepackage{newtxtext,newtxmath}

\usepackage[T1]{fontenc}

\DeclareRobustCommand{\VAN}[3]{#2}
\let\VANthebibliography\thebibliography
\def\thebibliography{\DeclareRobustCommand{\VAN}[3]{##3}\VANthebibliography}

\usepackage{graphicx}	
\usepackage{amsmath}	
\usepackage{tablefootnote}
\usepackage{longtable}
\usepackage[dvipsnames]{xcolor}
\usepackage{float}
\usepackage{appendix}

\usepackage{scalerel}
\usepackage{tikz}
\usetikzlibrary{svg.path}

\definecolor{orcidlogocol}{HTML}{A6CE39}
\tikzset{
  orcidlogo/.pic={
    \fill[orcidlogocol] svg{M256,128c0,70.7-57.3,128-128,128C57.3,256,0,198.7,0,128C0,57.3,57.3,0,128,0C198.7,0,256,57.3,256,128z};
    \fill[white] svg{M86.3,186.2H70.9V79.1h15.4v48.4V186.2z}
                 svg{M108.9,79.1h41.6c39.6,0,57,28.3,57,53.6c0,27.5-21.5,53.6-56.8,53.6h-41.8V79.1z M124.3,172.4h24.5c34.9,0,42.9-26.5,42.9-39.7c0-21.5-13.7-39.7-43.7-39.7h-23.7V172.4z}
                 svg{M88.7,56.8c0,5.5-4.5,10.1-10.1,10.1c-5.6,0-10.1-4.6-10.1-10.1c0-5.6,4.5-10.1,10.1-10.1C84.2,46.7,88.7,51.3,88.7,56.8z};
  }
}

\newcommand\orcidicon[1]{\href{https://orcid.org/#1}{\mbox{\scalerel*{
\begin{tikzpicture}[yscale=-1,transform shape]
\pic{orcidlogo};
\end{tikzpicture}
}{|}}}}

\title{An Enigmatic 380 kpc Long Linear Collimated Galactic Tail}

\author[Zaritsky et al.]{
\newauthor
Dennis Zaritsky$^{1,\orcidicon{0000-0002-5177-727X}}$\thanks{E-mail: dennis.zaritsky@gmail.com}
Jacob P. Crossett$^{2,\orcidicon{0000-0002-9810-1664}}$,
Yara L. Jaff\'e$^{2,\orcidicon{0000-0003-2150-1130}}$,
Richard Donnerstein$^{1,\orcidicon{
0000-0001-7618-8212}}$,
\newauthor
Ananthan Karunakaran$^{3,\orcidicon{0000-0001-8855-3635}}$,
Donghyeon J. Khim$^{1,\orcidicon{0000-0002-7013-4392}}$,
Ana C.C. Louren\c{c}o$^{2,4,\orcidicon{0000-0002-4393-7798}}$,
Kristine Spekkens$^{5,6,\orcidicon{0000-0002-0956-7949}}$,
\newauthor
Ming Sun$^{7,\orcidicon{0000-0001-5880-0703}}$,
Benedetta Vulcani$^{8,\orcidicon{0000-0003-0980-1499}}$\\
$^{1}$ Steward Observatory, University of Arizona, Tucson, AZ, 85721, USA\\
$^{2}$ Instituto de F\'isica y Astronom\'ia, Universidad de Valpara\'iso, Avda. Gran Breta\~na 1111, Casilla 5030, Valpara\'iso, Chile\\
$^3$ Instituto de Astrof\'isica de Andaluc\'ia (CSIC), Glorieta de la Astronom\'ia, 18008 Granada, Spain\\
$^4$ European Southern Observatory (ESO), Alonso de Cordova 3107, Santiago, Chile\\
$^5$ Department of Physics and Space Science, Royal Military College of Canada P.O. Box 17000, Station Forces Kingston, ON, K7K 7B4, Canada\\
$^6$ Department of Physics, Engineering Physics and Astronomy, Queen's University, Kingston, ON, K7L 3N6, Canada\\
$^7$ Department of Physics and Astronomy, University of Alabama in Huntsville, 301 Sparkman Dr NW, Huntsville, AL 35899, USA\\
$^8$ INAF- Osservatorio astronomico di Padova, Vicolo Osservatorio 5, I-35122 Padova, Italy\\
}

\date{}

\pubyear{2023}

\begin{document}
\label{firstpage}
\pagerange{\pageref{firstpage}--\pageref{lastpage}}
\maketitle

\begin{abstract}
We present an intriguing,  serendipitously-detected system consisting of an S0/a galaxy, which we refer to as the ``Kite", and a highly-collimated tail of gas and stars that extends over 380 kpc and contains pockets of star formation. 
In its length, narrowness, and linearity the Kite's tail is an extreme example relative to known tails.
The Kite (PGC 1000273) has a companion galaxy, Mrk 0926 (PGC 070409), which together comprise a binary galaxy system in which both galaxies host active galactic nuclei. Despite this systems being previously searched for signs of tidal interactions, the tail had not been discovered prior to our identification as part of the validation process of the SMUDGes survey for low surface brightness galaxies. We confirm the kinematic association between various H$\alpha$ knots along the tail, a small galaxy, and the Kite galaxy
using optical spectroscopy obtained with the Magellan telescope and measure a velocity gradient along the tail. The Kite shares characteristics common to those formed via ram pressure stripping (``jellyfish" galaxies) and formed via tidal interactions. However, both scenarios face significant challenges that we discuss,  leaving  open the question of how such an extreme tail formed. We propose that the tail resulted from a three-body interaction from which the lowest-mass galaxy was ejected at high velocity. 
\end{abstract}

\begin{keywords}
galaxies: kinematics and dynamics
galaxies: structure
galaxies: formation
galaxies: dwarf
\end{keywords}



\section{Introduction}
\label{sec:intro}

Galactic tails, or more broadly galactic detritus, may be a signature of a process or event acting to transform galaxies. As such, their discovery and characterisation help us unravel how galaxies evolve. 
The classic example of such an analysis is that of galactic tidal features \citep{toomre2,toomre}, which triggered a revolution in our understanding of the role of interactions and mergers of galaxies \citep[cf.,][]{schweizer,barnes}. Mergers and accretion are now central to the accepted hierarchical formation paradigm \citep[e.g.,][]{davis} and individual systems at various stages along a merger sequence have been identified \citep{hibbard}.

Gravity is not the only force that can extract matter from a galaxy. For example, galaxies moving through environments with a sufficiently high ambient density (such as the intracluster medium of massive clusters) can lose gas due to ram pressure stripping \citep{gunngott}.  
Among the most convincing examples of ram-pressure stripping at play are H{\small I} observations of cluster galaxies displaying gaseous tails and undisturbed stellar disks \citep{haynes,cayatte,chung,jaffe} and the  "jellyfish" galaxies seen in clusters \citep{kenney, Smith10, Ebeling14,Fumagalli14, Poggianti16, gasp, McPartland16, Jaffe18} and groups \citep{Vulcani21, Kolcu22}. 
It is expected that such hydrodynamic interactions, as well as other environmental effects, are responsible for the increased fraction of quenched galaxies in dense environments \citep[e.g.\ ][]{dressler,peng10}.

Extreme examples of galactic detritus garner attention because they challenge our quantitative understanding of these phenomena. For example, the recent discovery of a 250 kpc-long HI tail in the outskirts of the galaxy cluster Abell 1367 has proven difficult to explain in any scenario \citep{scott}. Sometimes, a particular example elicits bold new suggestions, such as that invoking a runaway supermassive black hole \citep{vandokkum}.

We report the discovery of an extraordinary tail with a variety of interesting features. First,  with a projected of 380~kpc (7~arcmin), it is the longest optical tail of which we are aware. 
This length is $>$ 6 times that of 
the longest jellyfish tails seen in H$\alpha$ \citep[e.g., D100 and JO206,][]{yagi,gasp}
and longer than most traced by H{\small I} emission
\citep[e.g., J0206 and FGC 1287;][]{rama,scott}, with the exception of a $\sim 500$ kpc long, amorphous feature tailing a Virgo cluster galaxy pair \citep{koopmann}. The only even longer galactic structures are some radio-detected, head-tail systems that can reach lengths $>$ 600 kpc \citep{vallee} but which are clearly associated with relativistic electron jets interacting with the intracluster medium.
Second, it is one sided, emanating along the disk plane of its apparent host galaxy. Third, it is highly collimated, with a length to width ratio of $\sim$ 40. Fourth, it is exceedingly close to linear in projection.
Fifth, it originates from an S0/a galaxy, which we would expect to be gas-poor, yet the tail is sufficiently gas-rich to support star formation along its length. Sixth, it lies in a low-density galactic environment with no cluster or group nearby, but is in a close binary galaxy system where both galaxies have active galactic nuclei. 

\begin{figure}
\centering
\includegraphics[width=\columnwidth]{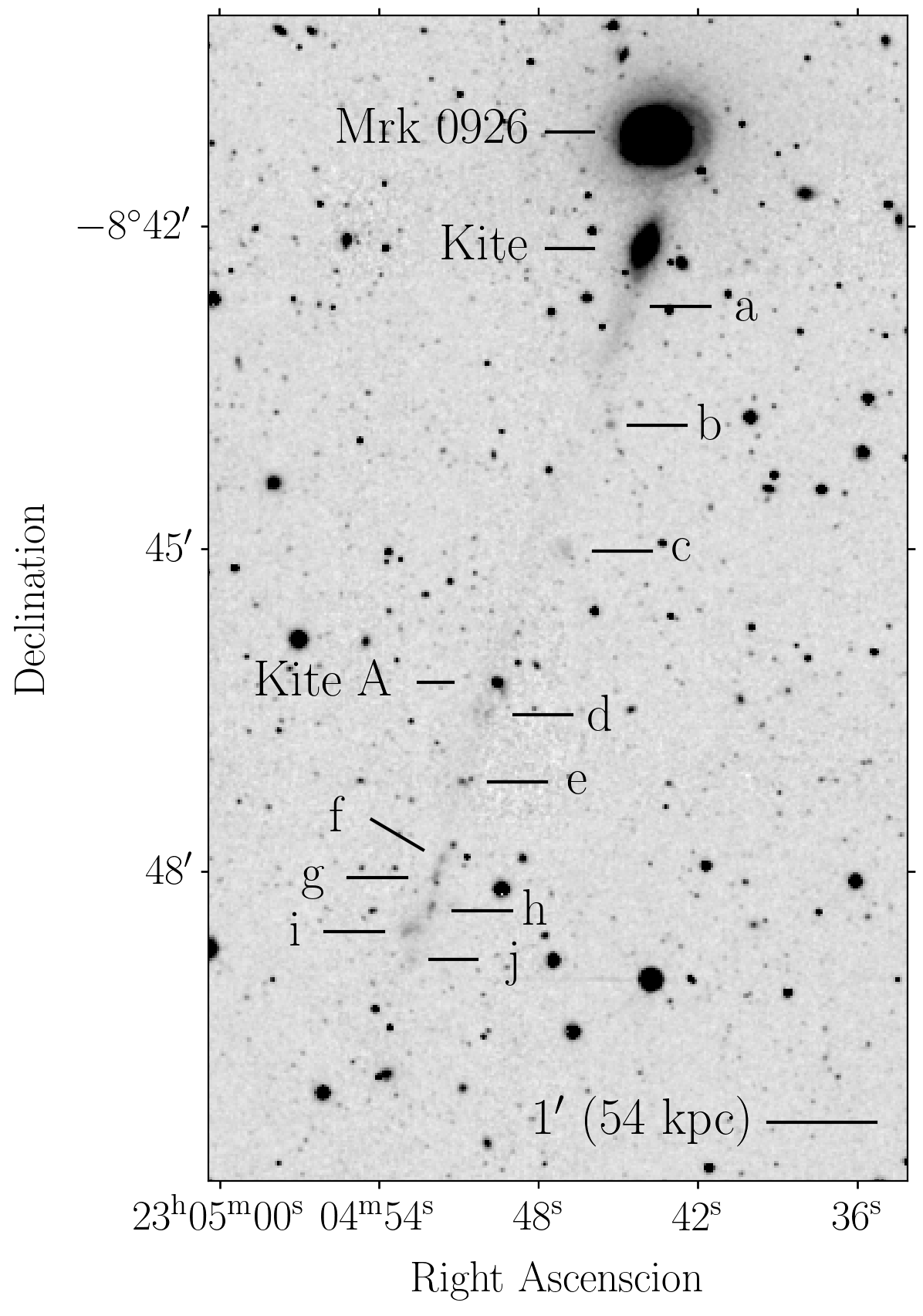}
\caption{The Kite galaxy and its companion, Mrk 0926. Features along the tail are labelled a-j as shown. The candidate disrupted galaxy, Kite A, is also labelled. Image obtained from the Legacy Survey public archive and is in the $g$-band. Bar at bottom right shows the length of 1 arcmin or the equivalent physical length at the distance of the Kite, 54 kpc.}
\label{fig:kite}
\end{figure}

Aside from inspiring questions regarding their origin and the impact of such phenomenon on galactic evolution,
tails and other galactic detritus are used to address a range of unrelated topics, including the nature of the dark matter potential \citep{dubinski,springel}, the formation of dark matter-free, tidal dwarf galaxies \citep{mirabel,duc,barnes_dwarfs,hunsberger,elmegreen}, the nature of star formation in an environment different than that typically found within galaxies \citep{degrijs,knierman,boquien,giunchi}, and the character of the circumgalactic and intergalactic environment \citep{sun07,tonnesen,fossati,vulcani,msun}. It stands to reason that extreme cases provide novel constraints for all of these topics to exploit.

The discovery and initial characterisation of the galaxy that we have named the ``Kite" on the basis of its morphology are described in \S\ref{sec:discovery}. Because of the possibility of chance superpositions of features enhancing the tail's appearance, length, or coherence, and to search for signs of current star formation, we obtained optical spectroscopy along the tail. The spectroscopic observations and results are presented in \S\ref{sec:spectro}. We present a brief discussion of various aspects of this system in \S\ref{sec:discussion}, highlighting areas that place the strongest constraints on possible formation scenarios and where subsequent investigation is warranted. We adopt a standard $\Lambda$CDM cosmology \citep[flat, $\Omega_m=0.282$, H$_0 = 69.7$ km s$^{-1}$ Mpc$^{-1}$;][]{hinshaw} when needed.

\section{Discovery and Characterisation}
\label{sec:discovery}

We serendipitously
discovered
the Kite during 
a search for low surface brightness (LSB) galaxies in the Legacy Survey \citep{dey} images \citep[Systematically Measuring Ultra-Diffuse Galaxies (SMUDGes);][]{smudges1,smudges2,smudges3,smudges5}. The list of candidate LSB galaxies returned by the search algorithm is contaminated by a number of artificial and physical sources that are not the intended targets. Briefly, SMUDGes processes the images by removing or replacing high surface brightness sources, filtering the residual images for angularly-large sources, proceeding through a variety of selection steps to winnow the number of LSB detections, and  ultimately producing a list of high-confidence ultra-diffuse galaxy  (UDG) candidates. The final step in the vetting process includes a visual examination of the candidate list. One of those objects is the subject of this study.

Upon visual inspection, we identified the candidate located at $(\alpha,\delta)= (346.2203^\circ,-8.8093^\circ)$, which is object i in Figure \ref{fig:kite}, to be near the end of a linear sequence of low surface brightness features leading back to PGC 1000273 $(346.183294^\circ,-8.703178^\circ)$, an edge-on S0/a galaxy \citep{nair} with $cz = 13813 \pm 25$ km s$^{-1}$ in the CMB frame and M$_i = -21.41$ AB mag (recessional velocity and magnitude from NED\footnote{The NASA/IPAC Extragalactic Database (NED) is funded by the National Aeronautics and Space Administration and operated by the California Institute of Technology.}) that we have named the ``Kite" galaxy given its single-sided tail morphology (Figure \ref{fig:kite}). 
It has a companion galaxy, projected only 57 kpc away at a comparable redshift $(z = 0.0470)$, that is a QSO (Mrk 0926, or alternately PGC 070409). 
This pair had attracted previous attention because  the Kite galaxy also shows  signs of nuclear activity \citep[identified spectroscopically as ``composite" by][in a study of binary active galaxies]{liu},
although neither \citet{liu} nor a subsequent study \citep{weston} identified clear signs of an interaction between the galaxies.

The tail structure comprises both well-defined knots of emission and diffuse emission strewn nearly linearly out to a projected separation from the Kite of 7 arcmin (420.0 arcsec from feature j in Figure \ref{fig:kite} to the center of the Kite galaxy). This separation corresponds to 380 kpc at the galaxy's adopted distance (187 Mpc) . The width of the tail is more difficult to constrain precisely although it is comparable to the width of the Kite galaxy itself, which is $\sim$ 10 arcsec. This suggests a length-to-width ratio for the tail of $\sim$ 40. The position angles on the sky between the Kite galaxy and the various letter-labelled features along the tail are presented in Table \ref{tab:results} and illustrate the near-linearity of the tail. Features b and c are somewhat less consistent with a linear distribution, but the dispersion in position angle even including these two is only 2.5$^\circ$ (it is 0.6$^\circ$ without those two features). Finally, a small galaxy, Kite A, is also projected onto the tail.

\begin{figure}
\centering
\includegraphics[scale=0.25]{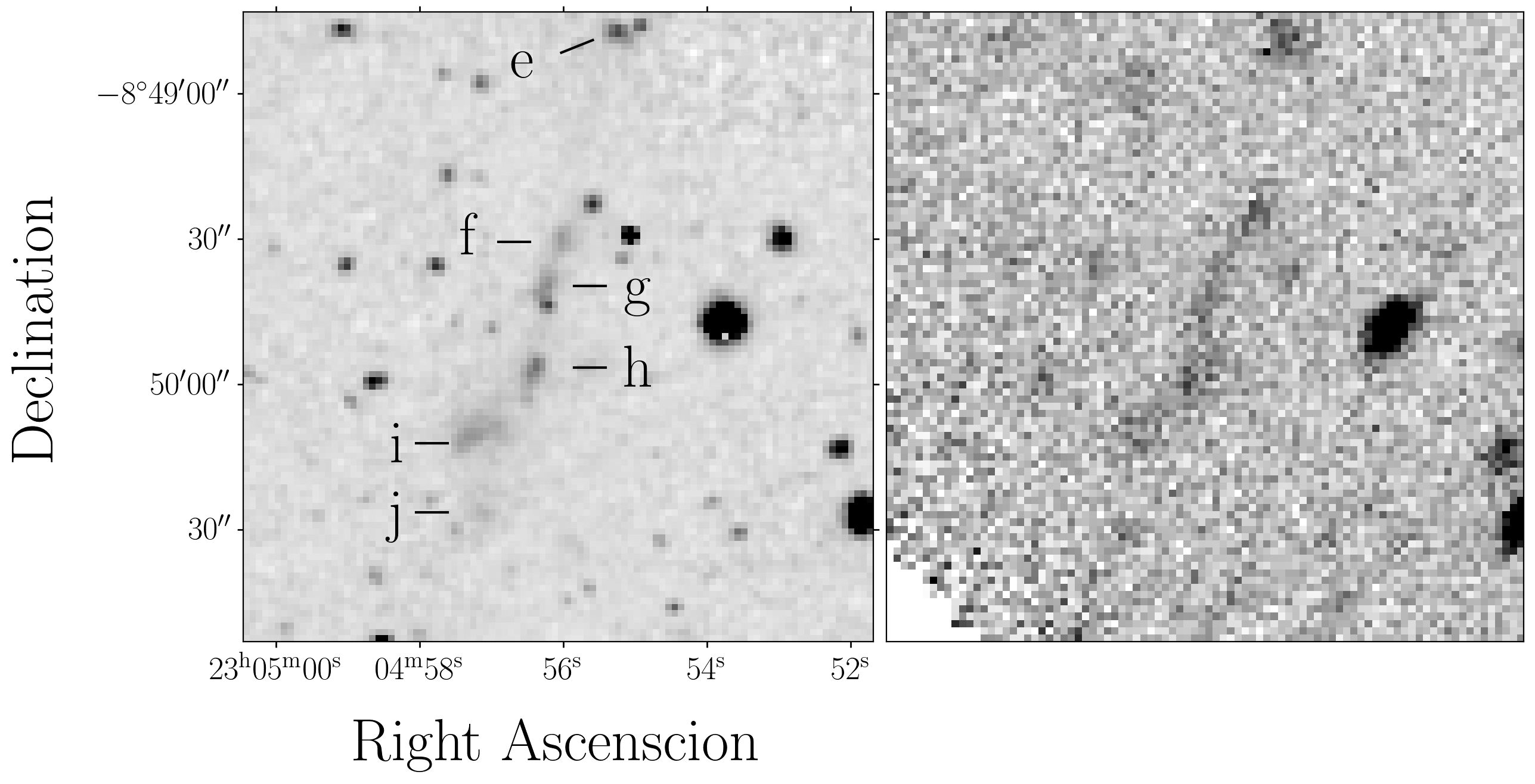}
\caption{Optical and NUV features toward the end of the Kite tail. Left panel shows a zoom-in view using the $g$-band image from Figure \ref{fig:kite} with features labelled. Right panel shows the same area drawn from a {\it GALEX} NUV image from the Medium Imaging Survey. }
\label{fig:kite_uv}
\end{figure}

Some of the features identified in Figure \ref{fig:kite} also have NUV emission (1771 - 2831 \AA) that is detectable in the {\sl GALEX} \citep{galex} Medium Imaging Survey \citep{bianchi14} images (see Table \ref{tab:results}). We illustrate this finding of knots near the end of the tail in Figure \ref{fig:kite_uv}. A check mark in the NUV column of Table \ref{tab:results} indicates that we visually identified emission in the {\sl GALEX} image corresponding to the location of the optically-identified associated feature. NUV emission typically indicates the presence of relatively young ($\lesssim$ few 100 Myr) stellar populations \citep[e.g.\ ][]{bianchi}.
The presence of such young stars helps differentiate these knots from other features in the image and aids in confirming that there are no similar LSB structures beyond feature j of the tail, on either side of what we have defined to be the tail, or on the opposite side of the Kite galaxy.

The local environment of the Kite and Mrk 0926 is not one conducive to ram pressure stripping.
The Kite and Mrk 0926 form a triplet system with the galaxy SDSS J230439.88-083712.6 \citep{Yang2007}. This third galaxy is 2.3 magnitudes fainter than the Kite galaxy and $\sim300$ kpc away, with no visible signs of debris or interaction with the main pair. Given its distance, morphology, and lower mass, this galaxy is unlikely to be related to this feature. Additionally,
there are no other known clusters or groups around this system within 3 Mpc and 4000 km s$^{-1}$ \citep{Yang2007, Szabo2011, Wen2012}. In fact, there are only three additional galaxies listed in SDSS DR7 within 1 Mpc of the Kite galaxy. With such a low local density, it  unlikely that there is a sufficiently dense medium to produce the observed tail via ram pressure stripping.

\section{Spectroscopic Follow-Up}
\label{sec:spectro}

\subsection{Tail Features}

We observed most of the features labelled in Figure \ref{fig:kite} using the Baade Magellan telescope and the IMACS spectrograph \citep{imacs} with the f/4 camera on the nights of 25 and 26 September 2022 in the long slit mode.
We selected the 1200 line grating, for a spectral dispersion of 0.2\AA\ pix$^{-1}$ (we then binned 2$\times 2$ pixels) and a wavelength coverage of 6000 to 7600\AA, although here we focus solely on the spectral region around H$\alpha$. We generally obtained three 15 min exposures, although in two slit positions we were limited to a single 15 minute exposure by observing conditions, which were generally variable. We reduced the data in a standard manner, extracted 1-D spectra, visually examined the spectra for emission and absorption lines, and measured the observed wavelength of the lines using a Gaussian line fitter. We present heliocentric measurements of $cz$ in Table \ref{tab:results} and use spectra with multiple emission lines to estimate that the internal recessional velocity uncertainty is 14 km s$^{-1}$. The comparison with the SDSS velocity for the Kite galaxy itself suggests a possible systematic error as large as $\sim 190$ km s$^{-1}$, but perhaps the SDSS value includes the emission line velocity we measure, which is much closer to the SDSS value. Regardless, when comparing our measured kinematics along the tail, the internal uncertainties are the ones of interest.

In all cases where we are able to measure a redshift for the targeted feature, that redshift measurement is consistent with the 
feature being at the distance of the Kite galaxy.
We typically find H$\alpha$ emission in the spectrum of each targeted feature, especially if we have the full 45 minutes of exposure time. However, because we employed offset pointing  based on broad band optical images, a lack of  H$\alpha$ emission for any target (denoted by the cross in Table \ref{tab:results}) does not necessarily imply a lack of H$\alpha$ flux in that feature. In two  features (a and e), we find measurable [N II] and [S II] in addition to H$\alpha$. These are weaker than H$\alpha$ and we do not incorporate them into the redshift measurement. In no other target than the Kite galaxy itself and Kite A do we detect enough continuum to attempt an absorption line measurement. With the exception of feature b, the measured  recessional velocities show a gradual, consistent decline from the velocity of the Kite galaxy toward the tip of the tail with a total amplitude of slightly over 150 km s$^{-1}$. 
The velocity gradient demonstrates that the tail is unlikely to be oriented exactly on the plane of the sky. Hence, it is likely  longer than 380 kpc. 

\begin{figure}
\centering
\includegraphics[scale=0.4]{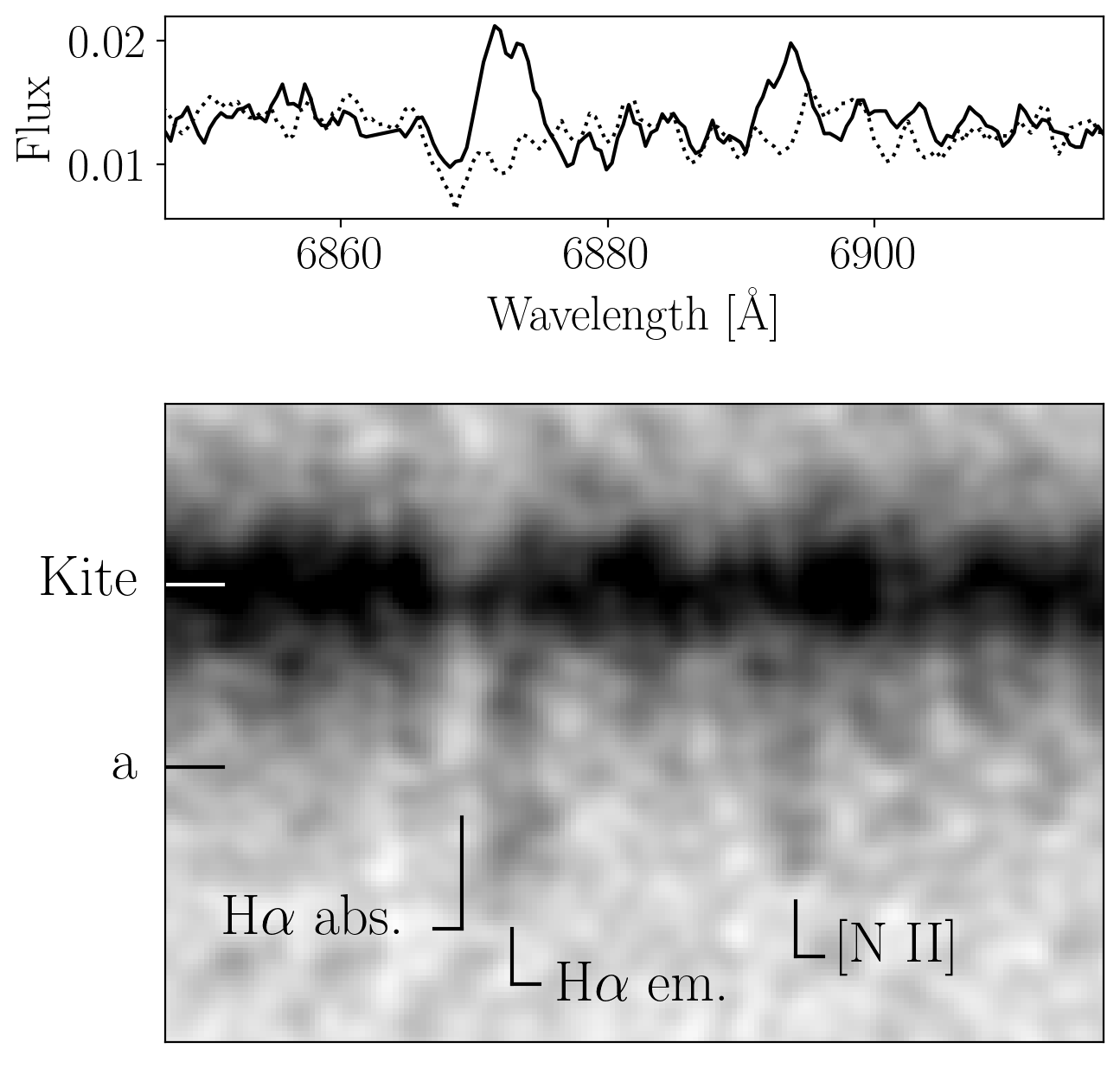}
\caption{Emission and absorption features for the Kite and the nearby Feature a from Figure~\ref{fig:kite}). The bottom panel shows a Gaussian smoothed zoom in on the Kite 2D spectrum. Spectral direction along the x axis and spatial along the y axis. Absorption and emission features labelled. The absorption affects the continuum symmetrically, while the emission is only present below the continuum spectrum. The H$\alpha$ emission is offset to larger $\lambda$ relative to the absorption. The shift corresponds to 160 km s$^{-1}$. The H$\alpha$ emission extends 12 arcsec below the continuum, which corresponds to 11 kpc. The upper panel shows extracted 1D spectra centred on the continuum (dotted line) and for the region below the continuum where H$\alpha$ emission is visible (solid line). We have normalised the continua of the two spectra to simplify comparison. H$\alpha$ absorption is seen in both, although deeper in the continuum spectrum, while H$\alpha$ and [N II] emission are clear in the off-continuum spectrum. Flux is in arbitrary linear units.}
\label{fig:kite_zoom}
\end{figure}

In Figure \ref{fig:kite_zoom} we highlight the kinematic behaviour of the tail near its origin (Feature a). We find that the ionised gas is kinematically offset from the stars in the galaxy, as traced by the stellar absorption features, even at small radii. The emission is also clearly asymmetric relative to the Kite even at this level of spatial resolution, with no convincing signature of a hidden second tail with different velocity. 

\subsection{Kite A}

Our spectrum of Kite A shows an asymmetric broad absorption line corresponding to H$\alpha$ in the observed frame of the Kite. Because of the asymmetry our determination of the line centroid is uncertain at the level of $\sim$ 100 km s$^{-1}$.  As such we can confidently place Kite A in the Kite environment, but cannot with similar confidence place it in the tail. Nevertheless, it is likely that Kite A is in the tail given its precise projection on the tail and reasonably close redshift. Our best estimate of the redshift (Table \ref{tab:results}) places it along the sequence of velocities for the other features. We speculate that the asymmetric absorption line arises because Kite A is rotating and the slit did not evenly sample both sides of the galaxy.

\begin{table}
\caption{Spectroscopic Follow-up Results}
\begin{tabular}{lcccccr} \hline  
Feature & $\alpha$ & $\delta$ & PA$^a$ & NUV & H$\alpha$ & $cz^{b}$ \\ 
&($^\circ$J2000)&($^\circ$J2000)&$(^\circ)$&&&(km s$^{-1})$\\
\hline
Kite$^c$ &346.1833 & $-$8.7032 &  ... & $\times$ & $\times$ & 13991 \\
a$^d$ & 346.1871 & $-$8.7157 &  163.3 & $\times$ & \checkmark & 14158 \\
b & 346.1885 & $-$8.7307 & 169.4 & \checkmark & \checkmark & 14328\\
c & 346.1954 & $-$8.7510 & 166.0 & \checkmark & ... & ... \\
Kite A&346.2062&$-$8.7709&162.0&\checkmark&$\times$& $\sim$ 14160\\
d & 346.2075 & $-$8.7753 & 161.6& \checkmark & \checkmark & 14109  \\
e$^e$ & 346.2116 & $-$8.7860 & 161.3 & \checkmark & \checkmark & 14058\\
f & 346.2151 & $-$8.7982 & 161.7 & \checkmark & $\times$ & ... \\
g & 346.2160 & $-$8.8009 & 161.7& \checkmark & ... & ...\\
h & 346.2166 & $-$8.8055 & 162.2& \checkmark & \checkmark & 13989 \\
i & 346.2201 & $-$8.8090 & 161.0& \checkmark & $\times$ & ...\\
j & 346.2196 & $-$8.8142 & 162.1& \checkmark & ... & ...\\
\hline
\end{tabular}
\noindent
$^a$ PA is measured on the sky, N to E with the Kite galaxy at the origin. No PA measurement is presented for the Kite because of isophote twisting.

\noindent
$^b$ Heliocentric corrected recessional velocity. The uncertainty is estimated to be 14 km s$^{-1}$, except for the Kite A measurement due to uncertainties in the slit position.

\noindent
$^c$ SDSS $cz$ = 14180 km s$^{-1}$.

\noindent
$^d$ [N II] also detected.

\noindent
$^e$ [N II] and [S II] also detected.

\label{tab:results}
\end{table}

\section{Discussion}
\label{sec:discussion}

A natural initial interpretation of the Kite, given its one-sided, linear tail is that it is the result of ram pressure stripping (RPS). There is at least one likely RPS feature that is nearly as long \citep[250 kpc;][]{scott} although those authors question that feature's RPS origin. Another RPS tail \citep[D100;][]{d100} is only 60 kpc long but shares many morphological similarities with the Kite, and yet a third is both long and appears in an edge-on galaxy \citep[GMP 2640;][]{Smith10,grishin}. D100 has a length-to-width ratio of 30, is linear, and has star formation along its length \citep{cramer}, although it is more of a polar than a planar feature relative to its host galaxy. \cite{yagi} struggled to explain the origin of D100, invoking possible confinement by the ambient intracluster medium to maintain the high degree of collimation. Subsequent work \citep{jachym, cramer}  advanced the RPS interpretation and invoked an array of possible mechanisms to maintain the tail's narrowness \citep[e.g.,][]{tonnesen10,roediger08}. However, unlike D100 or GMP 2640, the Kite is not in a dense environment. Hydrodynamic interactions in lower density environments also occur but show much more subtle features than that of the Kite \citep{vulcani2021}. We conclude that something other than RPS is responsible for the Kite's tail.
What does this mean for the inferred origin of the Kite galaxy with its long narrow tail, and possibly for similar systems like D100 that are attributed to RPS but face difficulties when modelled in detail?

If we abandon the RPS interpretation for the Kite, then the next most natural interpretation is that of tidal forces. The tail material could then come either from the Kite galaxy itself, as tidal material presumably drawn out by a close passage with Mrk 0926 or a third galaxy (presumably Kite A), or from a tidally shredded low mass satellite, whose remaining core may be Kite A. Because tidal tails tend to follow the elliptical nature of the original orbit, the projected linear nature of the tail suggests that in either scenario we are viewing the system at a particularly fortuitous orientation where the interaction happened on a plane perpendicular to the plane of the sky. This requirement may be somewhat more plausible in the Kite-Mrk 0926 interaction scenario because we at least know that we viewing the Kite galaxy itself nearly edge-on.

The scenario where the tail is the result of the near destruction of a satellite galaxy presents an avenue for solving the difficulty in finding sufficient gas in an S0/a galaxy to form a star-forming tail and the lack of obvious morphological disturbance in either the Kite galaxy or Mrk 0926. However, it faces challenges of its own. First, it requires the additional condition that the satellite orbit lies in (nearly) the same plane as the Kite's disk. Second, distributing material from roughly $r = 0$ to at least 380 kpc requires a large impulse difference, $\Delta E$, among the dwarf galaxy constituents that then implies a nearly direct collision with the Kite and a gravitational potential for the Kite that is sufficiently centrally concentrated that small differences in impact parameter translate to large $\Delta E$. Even so, much of the material ejected with the largest velocities (that currently at the tip of the tail) must be gaseous to power the ongoing star formation and would need to find its way through the disk plane of the galaxy. A candidate for the surviving core of this galaxy is Kite A, located near Feature d at $(346.2061^\circ,-8.7710^\circ)$.

Accepting geometrical coincidences, both scenarios face additional hurdles. A close interaction between the Kite and Mrk 0926, which would be required to generate a long tidal tail, would also tend to form a bridge between the galaxies and disturb the  morphology of the interacting galaxies \citep{toomre,barnes}, neither of which is observed. Additionally, the interaction typically generates a second long tail from the other galaxy, but none is observed to emanate from Mrk 0926. The news is not all negative for this scenario, as giant tidal tails are often found to have star formation near their ends \citep{mirabel91,mirabel}.

Regardless of what forces created the tail, there are some general puzzles that the tail poses. Consider that an age for the tail, $t$, can be estimated by dividing the length of the tail by the transverse velocity at which the tip of tail receded from the Kite galaxy, $v_t$. By doing so we estimate $t = 371/v_t$ Gyr, where $v_t$ is in units of km s$^{-1}$. We do not know $v_t$, but for typical values of galaxy internal velocities or pairwise velocity differences \citep{davis83} this implies lifetimes of one to several Gyr. A measurement of the star formation history of the tail would be one test of this age estimate, but we cannot do this for the emission line regions with our current data.

Given Kite A's plausible association with the tail, we attempt to use its spectral energy distribution (SED) to place constraints on the  most recent star formation, which we might plausibly connect to the age of the tail.
We measure aperture magnitudes, where the circular aperture is defined to include the W1 flux and is 10.5 arcsec in radius.
Our measurements for the FUV, NUV, $g$, $r$, $z$, W1, and W2 AB magnitudes are 21.7, 21.3, 18.9, 18.1, 17.5, 17.6, and 17.9, respectively.
A value of FUV-NUV = 0.4 mag  indicates the presence of a young ($\lesssim$ 300 Myr) stellar population and is nearly a dust-free indicator \citep[E$_{FUV-NUV}$ = 0.11 E$_{B-V}$;][]{bianchi}. We aimed for a more robust determination 
using PROSPECTOR \citep{prospector} to fit stellar population models to the SED, but found that the uncertainties in age, once a range of plausible star formation histories is allowed, is too large to provide a meaningful constraint on our hypothesis.

Adopting an age for the tail of $\gtrsim$ 1 Gyr leads to several apparent problems. First, star formation, as inferred from the H$\alpha$ detections, is occurring at various locations along the tail and must be producing at least some high mass stars to produce the necessary ionising radiation. Note that H$\alpha$ flux alone in a tail is not sufficient to indicate star formation \citep{boselli}. However, in our case the knot morphology and the ubiquitous NUV flux, indicative of young stars, suggests that here the H$\alpha$ is indeed related to ongoing star formation.
This star formation rate must either be maintained over the $\gtrsim$ Gyr lifetime, or something must have triggered star formation at nearly 400 kpc from the galaxy in the last few million years. This may not be a insurmountable challenge because, as we mentioned before, such star formation is often observed in giant tidal tails \citep{mirabel91,mirabel} and clumping of matter along the tails is reproduced in simulations \citep{barnes_dwarfs,elmegreen}. Similarly, star formation is often observed in jellyfish galaxies \citep{Vulcani2018,poggianti19}.

Second, the dispersion in position angles among the identified features within the Kite tail, excluding features b and c which appear slightly offset from the low surface brightness continuous tail, is $0.6^\circ$. At 380 kpc this offset corresponds to 4 kpc. 
For a $\gtrsim$Gyr lifetime this offset implies that velocities perpendicular to the tail but on the plane of the sky can not differ by more than $\sim$4 km s$^{-1}$, which is smaller than the typical velocity dispersion in galaxies with substantial gas reservoirs. 
This suggests either that there is a mechanism acting to actively maintain the collimation, as invoked by \cite{yagi} and \cite{jachym}
for D100, or the tail is much younger than we estimate, which in turn implies that the transverse velocity is much larger than the typical internal velocities in galaxies. The latter might lead one to consider more exotic models such as that of a runaway massive black hole \citep{vandokkum}, although we note that the black hole model would face its own challenges here. Consider that any circumgalactic gas parcels that it may have influenced to form stars would have a velocity consistent with the halo velocity dispersion ($\sim 100$ km s$^{-1}$) rather than with the internal velocity dispersion of a putative satellite galaxy. 
Such a large velocity  would cause the star formation clumps to drift away from the tail axis at a rate of 125 kpc/Gyr. For us to find the clumps within $\sim$10 kpc of the tail axis along the entire tail requires that all of these clumps formed within the last $\sim$ 100 Myr, which in turn implies a black hole speed of nearly 4000 km s$^{-1}$ if it is to reach a distance of 380 kpc. As large as this value seems, it is not beyond the realm of possibility \citep{campanelli,healy}, although statistically unlikely \citep{schnittman}.

Finally, the linearity of the feature is also a challenge considering the likely asymmetric nature of the gravitational potential at large radii. Torques on the matter in the tail would seem likely to produce a bend in the tail over a Gyr timescale even if the original geometry allowed for projection to create the illusion of linearity. Again, a much shorter tail lifetime would alleviate such concerns.

\subsection{Our Proposal}

Given the various reasons to favour a short lifetime for the tail and the presence of Kite A in the tail, we propose a gravitational origin for the tail arising from the ejection of Kite A from the Kite-Mrk 0926 system. A hyperbolic orbit, where Kite A has a velocity significantly larger than the escape speed, not only addresses issues related to the lifetime of the feature but relaxes the orientation constraints because Kite A and its associated detritus would be travelling on near linear trajectories in 3-D. There is still the coincidental alignment of the orbital plane with the Kite's disk plane but perhaps that is advantageous in realising an interaction among the three galaxies that allows Kite A to be ejected. For a time since pericenter passage of 300 Myr, consistent with the youngest stellar population in Kite A, Kite A would need to have a mean transverse velocity of 460 km s$^{-1}$ to reach its current projected position, which is almost certainly larger than the escape speed at its current position. The tidal material would both lead and trail Kite A, resulting in a nearly linear feature with Kite A in the middle. Finally, a close interaction, which is needed to provide a sufficient kick to Kite A, might also be responsible for fuelling the central black holes in the Kite and Mrk 0926 \citep{liu,weston}.

\subsection{Other Contexts}
We close the discussion by commenting on the implication of the existence of a system such as the Kite on other interesting systems. We identified this system initially because Feature i was flagged as a potential UDG. In any of the formation scenarios envisioned, clumps along the tail are expected to be dark matter free. If such clumps are able to remain gravitationally bound and survive they will contribute to a dark matter free galaxy population \citep{bennet,vdk18}. 
In fact, \cite{vdk22} note that their two dark matter free galaxies are part of a large linear sequence of galaxies. A potential difference between those two galaxies and the features within the Kite tail is that the former host globular clusters. We do not yet know if any of the Kite features host globular clusters. 

A second interesting class of system is that of the diffuse star-forming isolated stellar systems found in the Virgo cluster \citep{jones}, but which might also exist elsewhere. Those authors note that these are at least 140 kpc from any nearby potential parent and are young. They have relatively high metallicities, suggesting their gas comes from more massive galaxies,  which they interpret to mean a likely RPS origin. However, if the origin of the gas is as proposed here, from a galaxy like Kite A, then one would expect to measure a relatively high metallicity.

\section{Summary}

We present the discovery of an extraordinary tail emanating from what we have dubbed the Kite galaxy. The tail is unusual in its physical length (a projected length of 380 kpc), its collimation (it has a length to width ratio of 40), and its linearity (all of the detected knots along the tail scatter in position angle by less than 3$^\circ$). It is oriented parallel to the disk of the Kite galaxy. The Kite galaxy and its nearby companion, Mrk 0926, are both active galaxies, with Mrk 0926 being by far the more active. 

We present results from spectroscopy at various points along the tail. There is recent and ongoing star formation along the tail, as evidenced by NUV and H$\alpha$ flux. The velocities show a moderate velocity gradient along the tail and demonstrate that the various knots are physically associated with the tail. We identify a galaxy, Kite A, along the tail that has a velocity that is consistent with it lying in the tail. This galaxy has UV emission that indicates the presence of young stars but does not show evidence for ongoing star formation.

The two most commonly invoked origin scenarios for tail features, ram pressure or tidal stripping, face significant challenges that we discuss. Of the two, we have a preference for a tidal origin, but acknowledge the difficulties in making such a model work. Some of the difficulties are mitigated if the age of the tail is quite short, but this supposition leads to  more exotic formation models. We propose that a three-body encounter between the Kite, Mrk 0926, and Kite A resulted in the rapid ejection of Kite A. The resulting hyperbolic orbit explains the linearity of the debris field and the tail's narrowness. Detailed simulations are necessary to assess the viability of this proposal.
We briefly describe how such events may also help explain other puzzling observations.
The Kite system is a record-breaking, enigmatic source that presents a variety of interesting problems to resolve.

\section*{acknowledgments}
DZ acknowledges financial support from AST-2006785 and thanks both the Astronomy Department at Columbia University for their gracious welcome during his sabbatical and Greg Bryan for discussions about this object.
JPC acknowledges financial support from ANID through FONDECYT Postdoctorado Project 3210709. 
YLJ acknowledges financial support from ANID BASAL project No. FB210003.
MS acknowledges support from the NASA grant 80NSSC22K0353, and the USRA award 9\_0221, under NASA contract NNA17BF53C.
BV acknowledges support from the INAF Mini Grant 2022 “Tracing filaments through cosmic time”  (PI Vulcani). ACCL thanks for the financial support of the National Agency for Research and Development (ANID) / Scholarship Program / DOCTORADO BECAS CHILE/2019-21190049. KS acknowledges support from the Natural Sciences and Engineering and Research Council of Canada. 
This research has made use of the NASA/IPAC Extragalactic Database, which is funded by the National Aeronautics and Space Administration and operated by the California Institute of Technology.
This paper includes data gathered with the 6.5 meter Magellan Telescopes located at Las Campanas Observatory, Chile. .

\section*{Data Availability}

The optical imaging presented here is  publicly available through the Legacy Survey (https://datalab.noirlab.edu/ls/dataAccess.php) while the NUV data is available through the Mikulski Archive for Space Telescopes (MAST; https://science.nasa.gov/astrophysics/astrophysics-data-centers/multimission-archive-at-stsci-mast). We present the results from Magellan spectroscopy in Table \ref{tab:results} and  the 2-D spectra will be shared upon reasonable request to the corresponding author.



\bibliographystyle{mnras}
\bibliography{zaritsky.bib} 

\bsp
\label{lastpage}
\end{document}